\documentclass[prr,twocolumn]{revtex4-2}

\usepackage[pdftex]{graphicx}
\usepackage{amsmath}
\usepackage{amsfonts}
\usepackage{mathrsfs}
\usepackage[utf8]{inputenc}
\usepackage[usenames]{color}
\usepackage{siunitx}
\usepackage{bm}
\usepackage{hyperref}

\begin{document}
\title{Symmetry breaking of dipole orientations on Caspar-Klug lattices}
\author{Simon \v Copar}
\affiliation{Department of Physics, Faculty of Mathematics and Physics, University of Ljubljana, Ljubljana, Slovenia}
\email{simon.copar@fmf.uni-lj.si}
\author{An\v ze Bo\v zi\v c}
\affiliation{Department of Theoretical Physics, Jo\v{z}ef Stefan Institute, Ljubljana, Slovenia}
\date{\today}

\begin{abstract}
Anisotropic dipole-dipole interaction often plays a key role in biological, soft, and complex matter. For it to induce non-trivial order in the system, there must be additional repulsive interactions or external potentials involved that partially or completely fix the positions of the dipoles. These positions can often be represented as an underlying lattice on which dipole interaction induces orientational ordering of the particles. On lattices in the Euclidean plane, dipoles have been found to assume different ground state configurations depending on the lattice type, with a global ordering in the form of a macrovortex being observed in many cases. A similar macrovortex configuration of dipoles has recently been shown to be the sole ground state for dipoles positioned on spherical lattices based on solutions of the Thomson problem. At the same time, no symmetric configurations have been observed, even though the positional order of Thomson lattices exhibits a high degree of symmetry. Here, we show that a different choice of spherical lattices based on Caspar-Klug construction leads to ground states of dipoles with various degrees of symmetry, including the icosahedral symmetry of the underlying lattice. We analyze the stability of the highly symmetric metastable states, their symmetry breaking into subsymmetries of the icosahedral symmetry group, and present a phase diagram of symmetries with respect to lattice parameters. The observed relationship between positional order and dipole-induced symmetry breaking hints at ways of fine-tuning the structure of spherical assemblies and their design.
\end{abstract}

\maketitle

\section{Introduction}

Anisotropic interactions are a key feature in many self-assembling systems in biological, soft, and complex matter, and the capability to precisely control the orientation and spatial arrangement at the particle level can translate into the macroscopic properties of the assemblies~\cite{Thorkelsson2015,Bouju2018,Abelmann2020}. At close distances, interactions between particles are usually dominated by hard-core repulsion, frequently modeled by the Lennard-Jones potential or simple steric exclusion~\cite{Scala2007,Tillack2016}. The assembly is often further directed by anisotropic weak interactions such as hydrogen bonding, van der Waals interactions, $\pi$-$\pi$ stacking, dipole-dipole interactions, and metal coordination~\cite{Bishop2009,Bilbeisi2014}. If the particles are electrostatically charged, these interactions have to be taken into account as well, as they can play a significant role both during the assembly and for the stability of the resulting structure~\cite{Walker2010,Siber2012,LiS_PhysRevE96_2017,Chevreuil2018}. Anisotropy of an interaction is often described by virtue of multipole expansion, modeling the interaction as a superposition of monopoles, dipoles, quadrupoles, and less frequently also higher multipole contributions~\cite{Wormer1977,Gresh2007,ALB2013a,Lund2016,ALB2017a}. The simplest example of an anisotropic interaction which goes beyond the monopole moment is the unscreened interaction between point dipoles, which plays an important role in many particle assemblies~\cite{BaskinA_ACSNano6_2012,Gharbi2013,Hernandez2016,Morphew2017,Morphew2018,Chevreuil2018,Abelmann2020} and is known to affect their thermal fluctuations and stability, as was demonstrated on the example of crystalline membranes~\cite{MauriA_AnnalsofPhysics412_2020}.

If the dipole interaction is to induce non-trivial order in the system, it must involve either an external field, boundary, or additional repulsive interactions that partially or completely fix the positions of the particles. Consequently, in the assembled state, the positional order of interacting particles can often be considered to be fixed, with the dipole interaction determining the orientational order of the particles on the underlying lattice. The importance of the lattice is visible when dipoles are arranged in the Euclidean plane, as the type of their orientational order---which includes an antiferromagnetic state, a periodic tiling of closed vortex lines, and a macrovortex state---strongly depends on whether their positions are arranged on, for instance, honeycomb, kagome, square, or rhombic lattice~\cite{BelobrovP_ZhEkspTeorFiz88_1985,BrankovJ_PhysicaA144_1987,ZimmermanGO_PhysRevB37_1988,OliveE_PhysRevB58_1998,Prakash1990,Vedmedenko2008,BaskinA_ACSNano6_2012,Schumann2012,MaksymenkoM_PhysRevB91_2015,MessinaR_EPL110_2015,SchildknechtD_PhysRevB100_2019}. Much less is known about the orientational order of dipoles in three-dimensional assemblies~\cite{LuttingerJM_PhysRev70_1946,RomanoS_PhysRevB49_1994,Abelmann2020} and assemblies confined to the surface of a sphere~\cite{andraz2020}. The latter case is particularly interesting, as many particle assemblies in complex matter are spherical in shape (including liquid droplets, hard and soft colloidal particles, micelles, vesicles, some cells, and viral capsids)~\cite{Royall2013,Popko2012,Kim2017,Athanasopoulou2017,Baker1999,Zandi2020}, and interactions between their building blocks often result in structures with high symmetry of the positional order~\cite{ZandiR_ProcNatlAcadSci101_2004,Llorente2014,Guerra2018,PrestipinoS_PhysRevA99_2019}.

On a sphere, translational periodicity is ill-defined and a suitable spherical equivalent of lattices must be considered, leading to novel phenomena specific to the sphere. Arrangements of dipoles on spherical triangular lattices resulting as solutions to the Thomson problem~\cite{Wales2006} are thus far the only spherical system where their configurations have been systematically studied~\cite{andraz2020}. There, it has recently been shown that if no additional restrictions are imposed, dipoles positioned on spherical triangular lattices arrange themselves into a polar vortex state, even when the underlying lattice has a high symmetry. As we show in this work, a completely different and novel behavior is observed in dipole configurations arranged on Caspar-Klug (CK) spherical lattices~\cite{Siber2020}. These lattices are ubiquitous in spherical structures with icosahedral symmetry, such as viruses and virus-like-particles~\cite{BruinsmaRF_AnnuRevCondensMatterPhys6_2015,RochalSB_Nanoscale9_2017,Siber2020}, where the positional order of their building blocks differs from triangular close packing due to the different interactions involved in their assembly. These building blocks impose additional symmetries on the final assembled structure, and if they carry in-plane electrostatic polarization they will orient themselves differently compared to unrestricted dipole systems, and so the symmetry has to be explicitly taken into account.

We utilize the CK lattice construction to obtain positional order of (electrostatic) point dipoles and calculate the orientational ground states of dipole-dipole interaction. We show that the symmetry of the ground state depends on the positions of the dipoles within the fundamental domain of the lattice, and ranges anywhere from full icosahedral symmetry to completely asymmetric structures. Imposing a symmetry higher than the one of the ground state naturally enforces a state with a higher eletrostatic energy, which however in most cases remains metastable when the symmetry restriction is lifted. We calculate symmetry phase diagrams with respect to the underlying lattices and perform stability analysis for select lattices. With this, we show that by choosing a correct lattice, dipole pair interactions can be utilized to induce a desired rotational symmetry of the final structure, suggesting a mechanism for fine-tuning the self-assembly of spherical structures.

\section{Theoretical background}

\begin{figure}
  \centering
  \includegraphics[width=\linewidth]{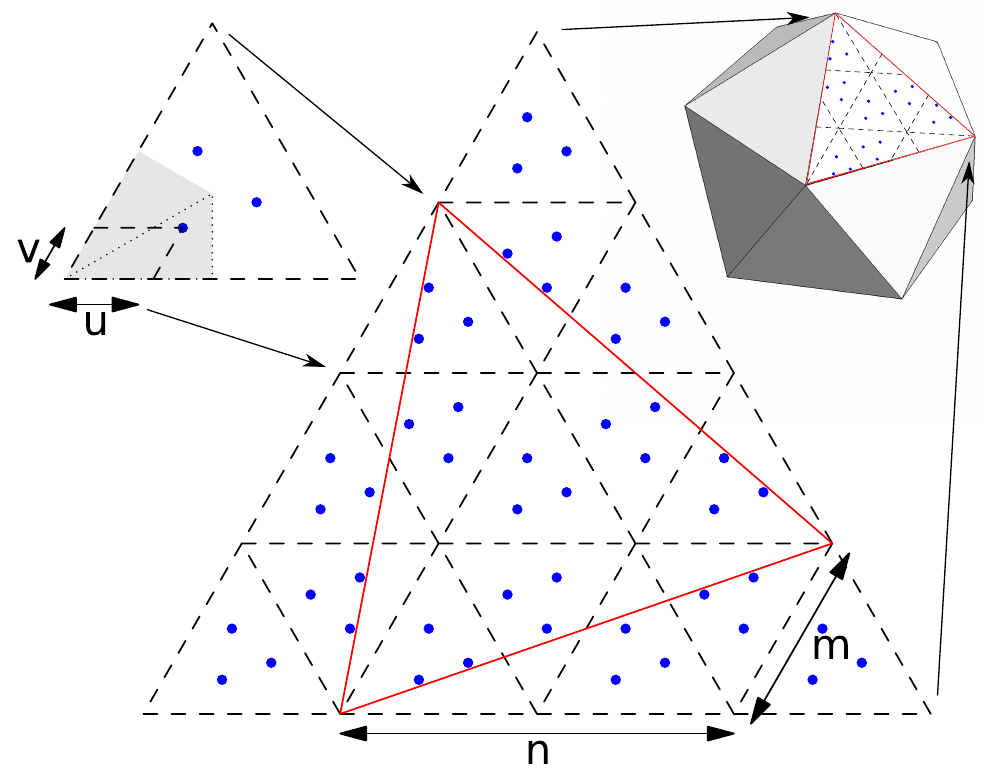}
  \caption{
    Construction of a CK lattice, where $(n,m)$ determines the triangulation number on the icosahedron, $T=n^2+nm+m^2$, with total number of dipoles given by $N=60T$. In this particular example, $n=2$, $m=1$, and thus $T=7$ and $N=420$. Placement of dipole positions (marked in blue) within the fundamental domain of the lattice (marked by the shaded deltoid in the unit triangle) is determined by the coordinates $(u,v)$. Reflecting the deltoid across its diagonal produces a mirror image of the same states, so we restrict our analysis to lattice positions in the lower right triangle.
  }
  \label{fig:ck}
\end{figure}

Having $N$ dipoles ${\bf p}_i$ positioned on the sphere at positions ${\bf r}_i$, we can state our problem in terms of minimization of the electrostatic potential energy, defined as the sum of dipole-dipole interaction energies over all pairs of dipoles:
\begin{equation}
  V=\sum_{i>j}V_{ij}=\sum_{i>j}\frac{3(\hat{\bf p}_i \hat{\bf r}_{ij})(\hat{\bf p}_j \hat{\bf r}_{ij})-\hat{\bf p}_i\hat{\bf p}_j}{||{\bf r}_{ij}||^3};
\end{equation}
${\bf r}_{ij}={\bf r}_i-{\bf r}_j$ are distances between dipoles, and unit vectors are denoted by a hat. We restrict the dipoles to lie tangentially to the sphere, as this has previously been shown to be the preferred solution for dipoles on a sphere in the absence of an external field~\cite{andraz2020}. We parameterize the dipoles relative to the local coordinate frame in the form $\hat{\bf p}_i=\hat{\bf e}^x_i\cos\phi_i +\hat{\bf e}^y_i\sin\phi_i$, allowing the problem to be restated in the matrix form (see Luttinger and Tisza~\cite{LuttingerJM_PhysRev70_1946}),
\begin{equation}
  V={\bf x}M{\bf x},\quad {\bf x}=\{\cos\phi_1,\sin\phi_1,\cos\phi_2,\ldots\}.
  \label{eq:matrix}
\end{equation}
Here, $M$ is a constant matrix that depends only on lattice geometry and the choice of local coordinate frames at each lattice point.

\begin{figure*}
  \centering
  \includegraphics[scale=0.65]{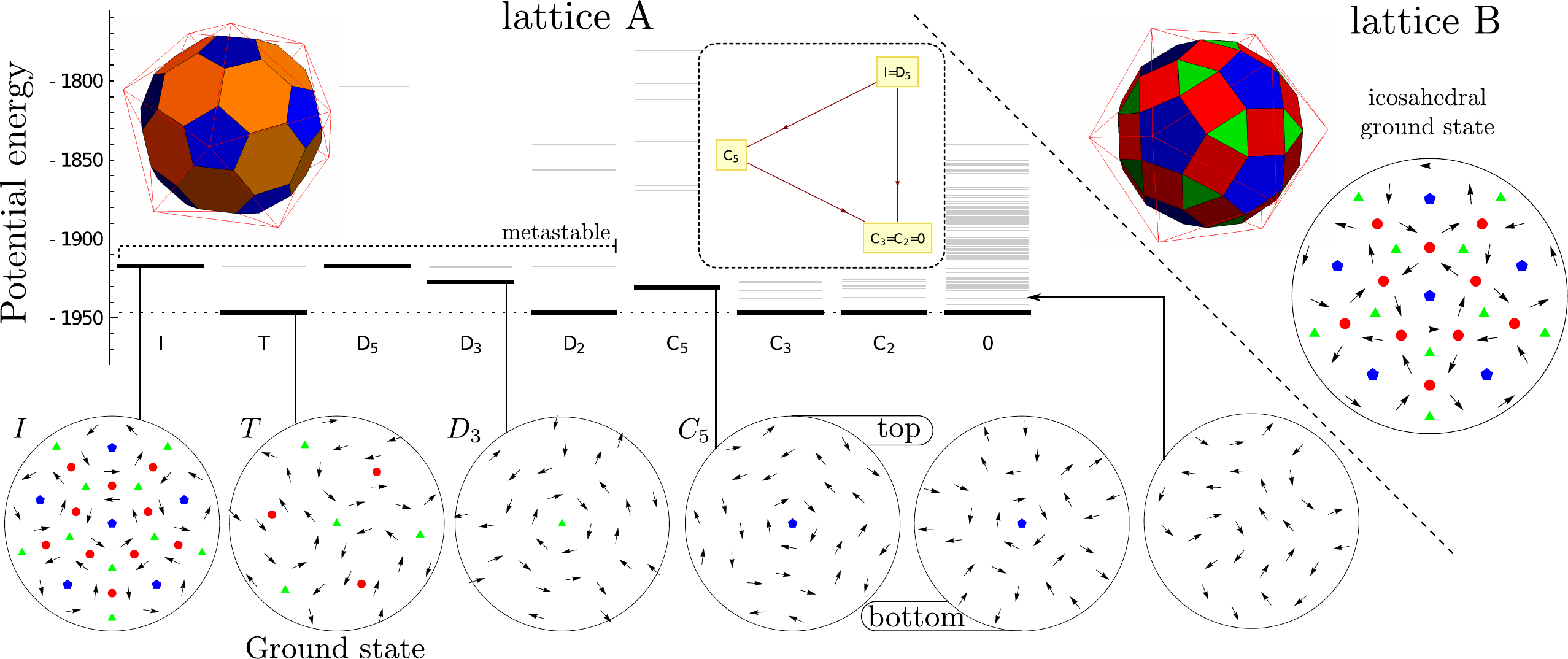}
  \caption{
    Symmetry-restricted lowest energy states for truncated icosahedron and small rhombicosidodecahedron (lattices A and B, respectively). On {\bf lattice A}, the ground state has tetrahedral symmetry ($T$). Ground states with lower symmetries are also present, and all of them---including the general higher energy local minima without symmetries---exhibit closed loops of dipoles arranged head-to-tail. Found local minima with higher energies are shown in gray, and they indicate that the icosahedral state remains metastable as long as dihedral symmetry ($D$) persists. Inset shows the possible symmetry breakings in which the previously lowest energy state becomes unstable and is no longer a minimum but a saddle point. The arrows show the available ways in which these transitions can proceed from higher to lower symmetry states. The ground state on {\bf lattice B} has full icosahedral symmetry ($I$) and consists of $5$-dipole loops around the symmetry axes. Dipoles are shown in the hemispherical azimuthal equidistant projection, and the symmetry axes are denoted by blue ($5$-fold), green ($3$-fold) and red ($2$-fold) markers.
  }
  \label{fig:spectrum-ab}
\end{figure*}

Dipole positions are fixed to a chosen CK lattice, which maps a triangular Euclidean tiling onto an icosahedron (see Fig.~\ref{fig:ck} for an example), with position vectors normalized to project them to the unit sphere. The positions are determined both by the CK parameters $(n,m)$ which determine the number of dipoles on the lattice and by the coordinates of the dipoles within the fundamental domain, parametrized by $(u,v)$ which correspond to unit vectors pointing from one vertex of the unit triangle to the other two. The number of lattice positions on a CK lattice, $N=60T$, is given by its triangulation number $T=n^2+nm+m^2$~\cite{Siber2020}. Unless stated differently, we limit our analysis to $(n,m)=(1,0)$, $T=1$ lattices with $N=60$ dipoles on the sphere.

To analyze the stability and transitions between states with different orientational symmetries, we first apply desired symmetry restrictions by equating angles that correspond to equivalent lattice points. Form in Eq.~\eqref{eq:matrix} is symmetry-reduced by adding together the corresponding rows and columns of matrix $M$, resulting in a smaller matrix \cite{LuttingerJM_PhysRev70_1946}. This is followed by minimization, performed by recursive application of gradient descent ${\bf x}\mapsto {\bf x}-\gamma M{\bf x}$ followed by renormalization of dipole vectors. Results are verified by comparison with Quasi-Newton method from Wolfram Mathematica~\cite{Mathematica}. Minimization is performed several hundred times to obtain both the ground state as well as the higher energy states with high certainty.

\section{Results and discussion}

\subsection{CK lattices of Archimedean polyhedra}

A single dipole in the fundamental domain of a $T=1$ CK lattice produces a tiling with $60$ dipoles, $5$ around each of the $12$ icosahedron vertices. The location of the dipole within the fundamental domain can be arbitrary---meaning that we can consider any combination of the dipole coordinates $(u,v)$ that falls into the fundamental domain---but three choices, which we will denote as lattices A, B, and C, are special as they lead to polyhedra with equal distances between the dipoles (Archimedean polyhedra). Lattice A with $(u,v)=(1/3,0)$ corresponds to a truncated icosahedron (a football), a spherical analog of the hexagonal tiling. Lattice B, $u=v=(3-\sqrt3)/6$, corresponds to a small rhombicosidodecahedron, a spherical analog of the rhombitrihexagonal tiling, and lattice C, $(u,v)=(2/7,1/7)$, corresponds to a snub dodecahedron, analogous to snub hexagonal tiling (also observed in viral capsids~\cite{RochalSB_Nanoscale8_2016}). Lattice C has a mirror image with $(u,v)=(1/7,2/7)$.

\begin{figure*}
  \centering
  \includegraphics[scale=0.65]{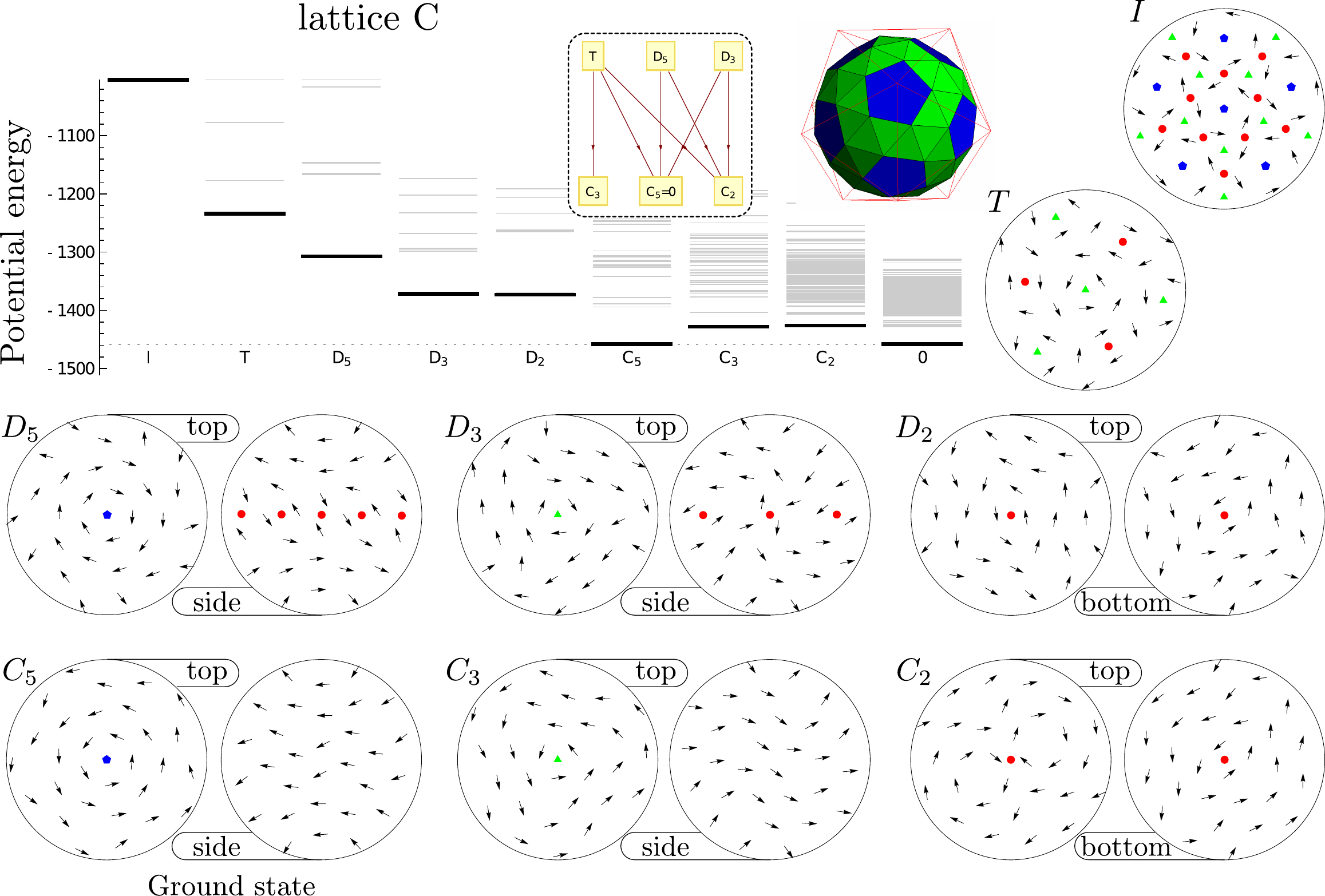}
  \caption{
    Symmetry-restricted lowest energy states for snub dodecahedron ({\bf lattice C}).
    The ground state has a $C_5$ symmetry. Dihedral symmetry $D_n$ forces a reversal of dipole circulation across the equator, while uniaxial rotational symmetry $C_n$ allows macrovortex states with much lower energies (best seen from the side view of $C_5$ and $C_3$ structures). Insets show that the lowest energy states of $T$, $D_5$ and $D_3$ structures are unstable with respect to certain symmetry breakings, while in the rest of the cases the lowest energy states remain local energy minima. The available transitions from higher to lower symmetry states are marked by the the arrows in the downward direction. Visualization follows the style from Fig.~\ref{fig:spectrum-ab}.    
  }
  \label{fig:spectrum-m}
\end{figure*}

The structure and behavior of the assemblies do not depend only on the ground state. Metastable states are often important, for example, if they are kinetically more accessible or if additional interactions or external stimuli trigger symmetry-breaking transitions between local energy minima. Figure~\ref{fig:spectrum-ab} shows the energy spectrum of dipoles on lattice A together with some of the corresponding dipole orientations, shown on a single hemisphere unless the distinction between the hemispheres is relevant to the discussion. The icosahedral ($I$) ground state at $V_I^{A}=-1917.13$ has a straightforward structure: it consists of closed loops of $5$ tail-chasing dipoles around each icosahedron vertex. The lowest energy state $V^{A}_{T}=-1946.67$ possesses tetrahedral ($T$) symmetry in which pairs of adjacent dipole loops merge into larger peanut-shaped loops. Two other symmetries have a unique lowest energy state: in the dihedral symmetry of degree three ($D_3$), a state at $V_{D3}^{A}=-1927.02$ is based on a $6$-dipole central loop and antiparallel neighboring dipole orientations, and in the uniaxial $5$-fold rotational symmetry ($C_5$), the state with $V_{C5}^{A}=-1930.61$ takes the form of antiparallel concentric loops with additional kinks on the opposite hemisphere. Importantly, none of these structures resemble the macrovortex state ubiquitously observed in dipoles arranged on Thomson lattices~\cite{andraz2020}, which is a direct consequence of the underlying CK positional order. A generic ground state without symmetry ($0$) consists of loops of dipoles meandering across the surface, as shown in one example in Fig.~\ref{fig:spectrum-ab}, and for each symmetry restriction, other local minima with higher electrostatic energies are found.

The lowest-energy structure with icosahedral symmetry remains metastable as long as the $2$-fold axis orthogonal to the main symmetry axis is present also in the state with the lower symmetry, such as in the tetrahedral and dihedral cases; otherwise it cascades into a lower energy state. In fact, most symmetry-restricted solutions remain local minima when symmetry is broken, except in a few select symmetry breakings: aforementioned transitions from $I$ to a group without a dihedral symmetry axis, and from $C_5$ if the $5$-fold symmetry axis is removed. Symmetry breakings that destabilize the local minimum are shown in the inset of Fig.~\ref{fig:spectrum-ab}. The most negative eigenvalue of the energy Hessian reveals the fastest decay mode when symmetry restrictions are lifted. The $I$ state decays via the same mode for any of the symmetries that make it unstable (inset of Fig.~\ref{fig:spectrum-ab}), while the decay rate of $C_5$ when all restrictions are removed is different---more unstable.

On lattice $B$, the icosahedrally symmetric state---consisting again of $5$-dipole loops around icosahedron vertices---is a global energy minimum, so any symmetry breaking has no effect (see Fig.~\ref{fig:spectrum-ab}). Lattice $C$ tells a different story (Fig.~\ref{fig:spectrum-m}): the ground state in this case has a $C_5$ symmetry and a macrovortex structure spanning across the entire hemisphere, similar to the ground states of dipoles on Thomson lattices~\cite{andraz2020}. This is expected, as a large part of the snub lattice consists of adjacent triangles, similar to those on closely packed spherical lattices. The structure with $C_3$ symmetry is similar, but has more distorted dipole loops, and the $C_2$ structure is also close to a macrovortex. Dihedral structures $D_5$, $D_3$, and $D_2$ have similar circumpolar structures, but the dihedral axes enforce antiparallel cycles on opposite hemispheres, leading to significantly higher energies. The only structures that decay upon symmetry breaking are the $D_5$, $T$, and $D_3$ structures, as shown in the inset of Fig.~\ref{fig:spectrum-m}.

\subsection{Symmetries and vector spherical harmonics}

Differences between configurations with different symmetries manifest themselves in their vector spherical harmonic (VSH) expansion. As the dipoles are restricted to lie tangent to the sphere, we can expand their configurations on any lattice over the orthonormal set of tangent basis vectors consisting of gradient (electric-type) $\bm{\nu}_{\ell m}^\textrm{g}=\sqrt{\ell(\ell+1)}^{-1}\, r\bm{\nabla}Y_{\ell m}$ and curl (magnetic-type) $\bm{\nu}_{\ell m}^\textrm{c}=\sqrt{\ell (\ell+1)}^{-1}\,\hat{\bm{r}}\times\bm{\nabla}Y_{\ell m}$ VSH (for details, see Refs.~\cite{ThorneKS_RevModPhys52_1980} and~\cite{Carrascal1991}). This further allows us to write the vector analog of the spherical structure factor in the form:
\begin{equation}
  S_{\ell}^{\text{g},\text{c}}=\frac1N\sum_{m=-\ell}^\ell\frac{4\pi}{2\ell+1}\left|\sum_{i=1}^N \bm{p}_i \cdot \bm{\nu}_{\ell m}^{\textrm{g},\textrm{c}}\right|^2.
\end{equation}
This definition follows the definition of the spherical structure factor for the standard (scalar) multipole expansion, and is trivially related to multipole magnitudes~\cite{ALB2019,Franzini2018}. In this way, it provides an insight into the nature of dipole structures and their symmetries.

\begin{figure}
  \centering
  \includegraphics[width=\linewidth]{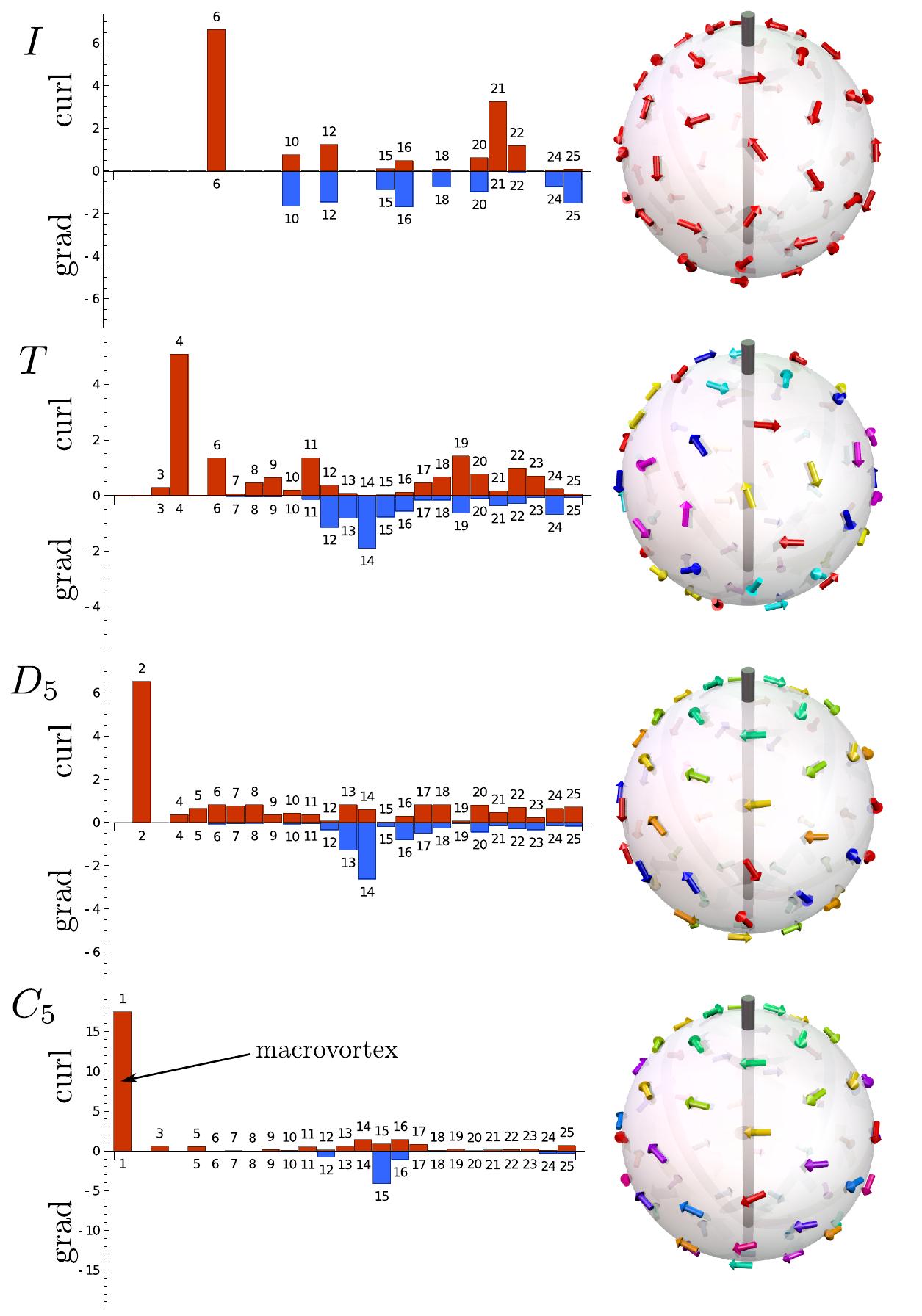}
  \caption{
    Curl and gradient VSH spectra, $S_\ell^{\text{c}}$ and $S_\ell^{\text{g}}$, for dipoles on lattice $C$ and with $I$, $T$, $D_5$, and $C_5$ symmetries, shown with respect to the spherical wave vector $\ell$ (orbital angular momentum number). Top histograms (red) show the contributions of curl VSH, and the bottom histograms (blue) show the contribution of gradient VSH. Depending on the symmetry of the structure, certain $\ell$ are forbidden and the corresponding coefficients are zero. Right side of the figure shows dipole configurations in 3D, where dipoles equivalent under symmetry operations have the same color. Vertical axis in these plots is one of the symmetry axes of the system ($3$-fold for the tetrahedral structure and $5$-fold for the rest). The ground state with $C_5$ symmetry is a macrovortex state, which is reflected in a large nonzero curl component $S_\ell^\textrm{c}$ at $\ell=1$.
  }
  \label{fig:vsh}
\end{figure}

Symmetries of different dipole configurations result in the restriction of the allowed spherical wave number $\ell$: for icosahedral symmetry, only the values of $\ell = 6i + 10j (+15)$ are permitted~\citep{ALB2013a}; for tetrahedral symmetry, $\ell = 4i+6j (+3)$ (only excluding $\ell=1$, $2$, and $5$); $D_5$ symmetry forbids $\ell=1$ and $\ell=3$ and $D_{2,3}$ forbid $\ell=1$. Figure~\ref{fig:vsh} shows the spectra of the vector spherical structure factor for dipoles on lattice $C$ with four different symmetries (also shown in Fig.~\ref{fig:spectrum-m}). We can see that they indeed observe the $\ell$ selection rules pertaining to each individual symmetry. Large components of the curl harmonics $S_{\ell}^\textrm{c}$ describe vortices (closed dipole loops) of different sizes. Specifically, the $\ell=1$ curl harmonic describes a macrovortex around a single axis, such as those seen in the solutions of $C_3$ and $C_5$ symmetries (the latter also being the ground state on lattice $C$) and on Thomson lattices~\cite{andraz2020}. Gradient terms $S_{\ell}^\textrm{g}$ describe alignment that resembles potential flow, and are less prominent here, because closed dipole loops are favored. In physical systems, gradient terms will be larger when localized negative and positive charges induce an additional potential field.  This analysis also shows that an approach using VSH is suitable for the analysis of empirical and simulation data.

\begin{figure*}
  \centering
  \includegraphics[width=\textwidth]{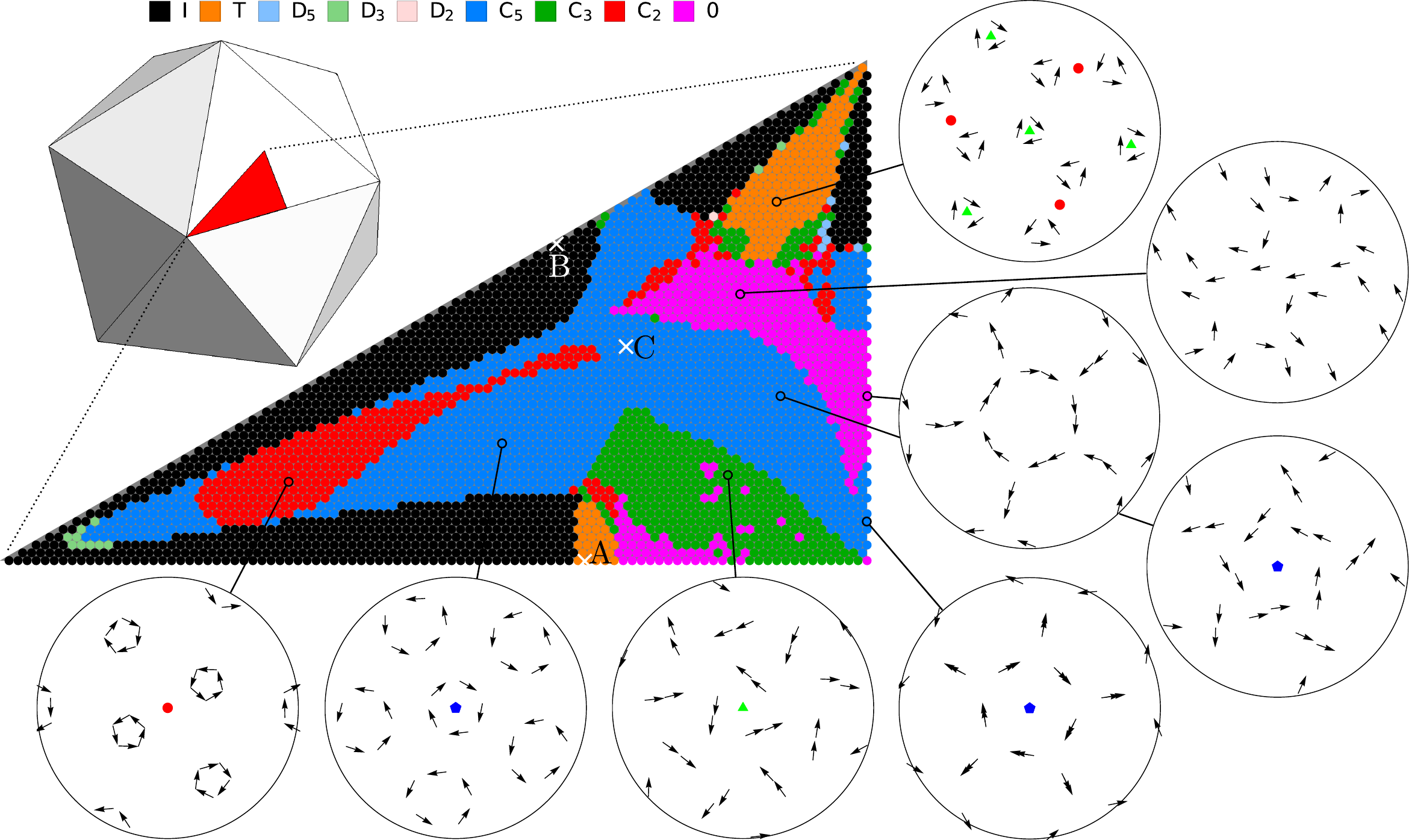}
  \caption{
    Phase diagram of ground state symmetries for a $T=1$ CK lattice with respect to the position $(u,v)$ of the dipole in the fundamental domain. Top views of select structures are shown in insets. Structures in the left part of the diagram are very similar to the icosahedral one, with barely noticeable symmetry breaking.  Bottom right side of the diagram features parallel coupling of neighboring dipoles. Note the asymmetric states at the right edge of the phase diagram, where symmetry is broken by buckling and reversal of direction of some of the dipole pairs. Isolated points of different symmetries, present throughout the diagram, can be attributed to numerical artefacts. Lattices $A$, $B$, and $C$, corresponding to the Archimedean solids shown in previous figures, are marked with crosses.
  }
  \label{fig:phase}
\end{figure*}

\subsection{General CK lattices with $T=1$}

Despite lattices A, B, and C being Archimedean polyhedra and thus all having constant nearest neighbor distances, their ground state symmetries and transitions between them are very different. The next question is how different lattices are related and how ground states and transitions look for general $(u,v)$ coordinates. We performed a ground state calculation for 3104 different dipole positions $(u,v)$ in the fundamental domain, and summarized the results in the symmetry phase diagram in Fig.~\ref{fig:phase}. We observe a very complex phase diagram that is based on the competition of different phenomena. In the left corner of the fundamental domain, the dominant interaction is between $5$ dipoles around the icosahedron vertices. These ground states consist of dipole loops which resemble those seen in $I$ structures, but their senses of rotation may alternate in different ways, giving rise not only to structures with $I$ symmetry (black markers in Fig.~\ref{fig:phase}), but also to $C_2$ and $C_5$ symmetries and even to $D_3$ symmetry in a very small portion of the phase diagram. In the top right corner of the fundamental domain, the proximity of $3$ dipoles dominates the interactions. This stabilizes the $T$ symmetry, and to a lesser extent, lower $3$-fold symmetries. Lower right part of the phase diagram puts dipoles into close pairs centered around the \emph{edges} of the icosahedra. These pairs tend to align and act as a single dipole, which, due to the polar nature of the dipoles, breaks all the $2$-fold symmetry axes a structure could have, so the main symmetries observed are $C_3$ and $C_5$. The middle of the phase diagram corresponds to states with balanced interactions between the closest neighbors, similar to lattice $C$ that is representative of this region. These lattices are locally triangular and feature macrovortex-like states.

Between these regimes, we observe a complex interplay of symmetries caused by competition between interactions that favor different structures. A large region of the phase diagram has no symmetries at all, and a snapshot corresponding to its right edge shows why: the resulting structure is similar to the $C_5$ structure with aligned dipoles, but some of the pairs are reversed and the pairs of dipoles are just far enough apart to allow ``buckling'' instead of acting as a single dipole. Other disordered structures are observed at the transition from tetrahedral to $5$-fold parts of the phase diagram. The energy landscapes of the observed configurations have many local minima, and even after many repetitions the lowest one is not always found. In the parts of the phase diagram where energies of states with different symmetries are close together, this leads to isolated points with incorrectly determined ground state symmetry.

\begin{figure}
  \centering
  \includegraphics[width=\linewidth]{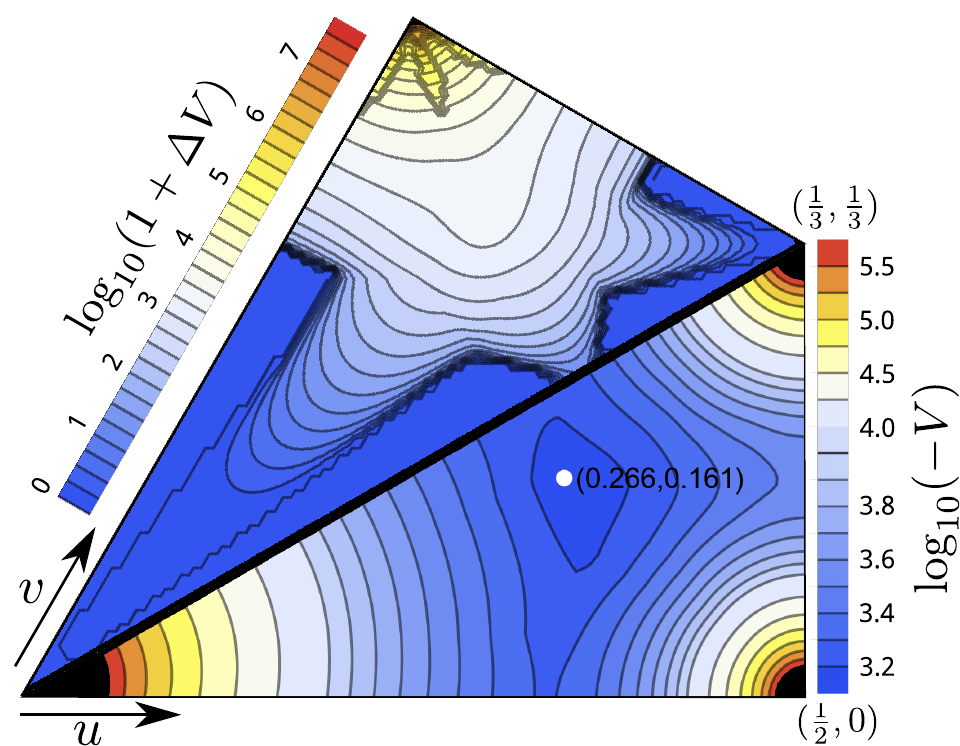}
  \caption{
    Electrostatic interaction energy in the ground state $V$ (lower right triangle) and difference between energies in the icosahedral and ground states $\Delta V$ (upper left triangle). The largest differences between states occur in the corner corresponding to the dipole pairing $(u,v)=(1/2,0)$, while the other two corners are compatible with $I$ symmetry, and the $C_5$ and $T$ regions only extend out as minor improvements in energy. In the energy plot, the position and coordinates of the absolute electrostatic energy maximum (minimum by absolute value, $|V|=1425.7$) are marked; it is located very close to the lattice $C$ coordinates $(2/7,1/7)$. For comparison, $60$ dipoles placed on a lattice derived from the Thomson problem have the energy $|V|=1378.0$~\cite{andraz2020}.
  }
  \label{fig:energy}
\end{figure}

Electrostatic energy of these systems is dominated by the closest neighbors due to the divergent nature of the dipole-dipole interaction. In Fig.~\ref{fig:energy}, we show both the total energy of the ground state with respect to the lattice parameters $(u,v)$ as well as the energy difference between the highest symmetry ($I$) state and the ground state. We see that $C_5$ symmetry offers an incremental improvement over $I$ structures on the angle bisector extending from the left corner of the fundamental domain, and the same holds true for the $T$ symmetry extending from the upper corner of the domain related to the $C_3$ symmetry axis. Conversely, the part of the phase diagram corresponding to dipole pairing offers significant improvement over the $I$ structure due to the very strong binding of aligned dipole pairs. The electrostatic energy is lowest by absolute value, $|V|_{\text{min}}=1425.7$, when the dipoles are farthest apart, which is very close to the lattice $C$. However, this is still higher than closely packed spherical lattices, such as the Thomson lattice, whose energy is $|V|_{\text{Th}}=1378.0$ for the same number of particles $N=60$.

\subsection{General CK lattices with $T=3$}

\begin{figure*}
  \centering
  \includegraphics[width=\textwidth]{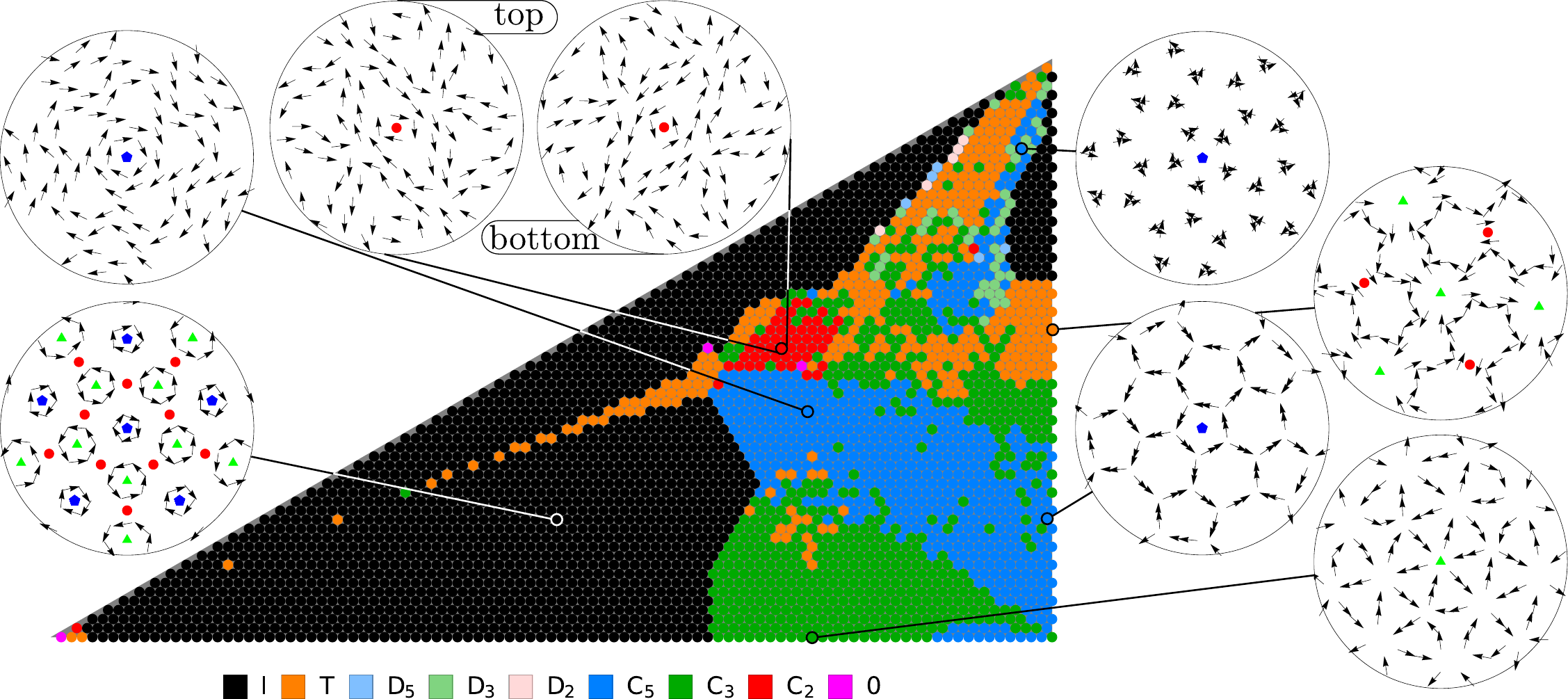}
  \caption{
    Phase diagram of ground state symmetries for a $(1,1)$, $T=3$ CK lattice with $N=180$ dipoles on the sphere. The rough division of the phase diagram is similar to the $(1,0)$ lattice (Fig.~\ref{fig:phase}), but with a much larger region of stability of icosahedral structures, and without any asymmetric ground states. Note that the lowest symmetry, $C_2$, is also restricted to only a very small portion of the phase diagram. More numerical artifacts are present compared to Fig.~\ref{fig:phase} due to a larger number of local minima that make the identification of the true ground state difficult.
  }
  \label{fig:phase2}
\end{figure*}

Results presented thus far were obtained for CK lattices with $T=1$, consisting of $N=60$ dipoles. In Fig.~\ref{fig:phase2}, we present a symmetry phase diagram equivalent to the one in Fig.~\ref{fig:phase} for a larger lattice with CK parameters $n=m=1$. This lattice has a triangulation number $T=3$ and thus contains $N=180$ dipoles. It features local hexagonal regions in addition to $12$ pentagons of the icosahedron vertices. The main observation on the $T=3$ lattice is that, in general, states with higher symmetries are preferred---icosahedral symmetry dominates almost the entire left portion of the phase diagram, and tetrahedral structures are observed in its upper right part. We observe no completely asymmetric ground states, and the region of $C_2$ symmetry is shrunk to a small patch in the middle of the diagram, with structures that can be described as longer strings of head-to-tail arranged dipoles. It is noteworthy that the ``fundamental domain'' of CK lattices with larger triangulation numbers $T>1$ is no longer the fundamental domain of the icosahedral symmetry group. Because of the $5$-fold lattice defects, the lattice sites are similar, but not equivalent. Not only do the dipoles have slightly different environments, but, more importantly, the dipoles around a hexagonal face can arrange in an alternating fashion while the pentagonal dipoles do not have that option. The ``almost symmetry'' between lattice sites is the most apparent in the left and upper corners of the diagram, where trimers and pentamers behave almost as independent entities, as seen in Fig.~\ref{fig:phase2}.

We can also expand dipole configurations on $T=3$ CK lattices in terms of VSH $\bm{\nu}_{lm}^\mathrm{g}$ and $\bm{\nu}_{lm}^\mathrm{c}$, where we again observe that different symmetries give rise to spectra of select wave vectors $\ell$ only. What is more, since 180 dipoles are positioned on a $T=3$ lattice, the spectra do not always peak at the lowest allowed $\ell$ (as was the case for $T=1$ lattice containing 60 dipoles, shown in Fig.~\ref{fig:vsh})---one can, for instance, observe spectra of icosahedrally symmetric lattices with peaks either at $\ell=6$, similar to a $T=1$ lattice, or at $\ell=10$ In this way, VSH expansion can be used to distinguish between different types of dipole order which otherwise possesses the same symmetry. For general $T$, the dominant $\ell$ scales inversely with the distance between nearest neighbors and is proportional to $\sqrt{T}$. For the same reason, for $(u,v)$ closer to the edge of the fundamental domain, higher spatial frequencies (higher $\ell$) will be present compared to the Archimedean lattices A, B, and C. High curl coefficient at $\ell=1$, equivalent to the angular velocity parameter introduced in Ref.~\cite{andraz2020}, is expected to signify the macrovortex state at any triangulation number.

The macrovortex state---the main type of ordering in the Euclidean space and on Thomson lattices on the sphere---is in general not energetically preferred on CK lattices, as they have a honeycomb-like structure instead of a closely-packed triangular one. The fundamental difference between CK lattices (honeycomb-like) and closely-packed lattices (triangular-like) are the $6$- and $5$-fold lattice vacancies that accommodate microvortex states---local dipole loops with no net dipole moment. With growing lattice size, ever larger parts of the lattice resemble the Euclidean honeycomb lattice and thus the effect of $5$-fold lattice defects and long-range interactions through the bulk of the sphere become less and less pronounced. This in turn leads to degeneracy, as loops with zero lowest order multipole moments interact very weakly across large distances over the sphere. With further increase in triangulation numbers, we expect the limiting regimes to persist: $3$-dipole microvortices (supporting $I$ or $T$ symmetries) in the top right part of the phase diagram, $5$-fold microvortices in the left part of the phase diagram, dipole pairings with $C_{3,5}$ symmetries, and most likely, macrovortex-like states in the middle of the fundamental domain, where triangular patches can be found on the lattice.

\section{Conclusions}

In contrast to the Euclidean case where lattices possess only translational symmetries, spherical lattices reflect the rich structure of the point symmetry groups in three dimensions. Nonetheless, this does not necessarily reflect in the orientational order of dipoles positioned on spherical lattices, as triangular lattices---based on solutions of the Thomson problem---lead to a single ground state in the form of a macrovortex, regardless of the underlying positional symmetry. Here, we have shown that when dipoles are positioned on spherical CK lattices instead, dipole-dipole interactions produce very diverse results. We have demonstrated that dipole pair interactions can conspire to stabilize any point symmetry, starting with the highest icosahedral symmetry of the CK positional order; however, dihedral symmetry is less favored than others. Fixed-position dipole order alone can therefore be used to control the orientational symmetry of the resulting structures. Furthermore, if the interaction can be varied, for example with screening, symmetry-changing transitions are possible. Symmetry phase diagrams also show how controlling the positions of the dipoles within the fundamental unit of the lattice can regulate the resulting symmetry of the structure and its stability. A drawback of this mechanism is the multitude of metastable states, which decreases the likelihood of finding the true ground state, although in potential experimental realizations, favored kinetic pathways could improve their reproducibility. If the structure is flexible, the symmetry can also reflect in deformations of the entire assembly, allowing shape control.

Our work also aims to stimulate the design and study of novel spherical assemblies where dipole moment would play a major role, and we show how the stability and structural transitions between such assemblies can be regulated using the dipole interaction. Spherical assemblies of dipoles could be designed experimentally by, for instance, Pickering emulsions of magnetic nanoparticles~\cite{Chevalier2013,Melle2005} or even with dye molecules around a nanosphere~\cite{Auguie2019,Tang2018}. Adsorption of gases on charged fullerenes also involves ordering of anisotropically charged molecules on a regular spherical lattice~\cite{RalserS_PhysChemChemPhys18_2016}. The general properties observed in our simplified model are expected to hold even for more complicated and modified cases---for positional lattices with different symmetries, such as octahedral or tetrahedral, and for dipole positions displaced radially from a perfect sphere, which is expected to be the case in biological systems where the structures themselves are polyhedral. Here, vector spherical harmonics present a natural way of analyzing such configurations and determining their symmetries. In a manner similar to the one presented in this work, it is also worth to explore other pair interactions such as the quadrupole-quadrupole interaction, which pertains to physical building blocks with head-tail symmetry and thus without polar order. Finally, the question of the ground state symmetry of ideal multipoles is also interesting from a purely mathematical perspective, just like the Thomson problem, which still inspires new discoveries even a century after its conception.

\begin{acknowledgments}
  Authors acknowledge support from Slovenian Research Agency (ARRS) under contracts P1-0099 (S\v C), P1-0055 (AB) and J1-9149 (S\v C, AB). This work is associated with the COST Action EUTOPIA (Grant No. CA17139).
\end{acknowledgments}

\bibliographystyle{apsrev4-1}
\bibliography{ckdipoles}

\end{document}